\documentclass{article}
\usepackage[T1]{fontenc} 
\usepackage[utf8]{inputenc} 
\usepackage{ismir,amsmath,cite,url}
\usepackage{graphicx}
\usepackage{amsfonts}
\usepackage{booktabs}
\usepackage{color}
\usepackage{multirow}
\usepackage{verbatim}
\usepackage{tikz}

\usepackage{lineno}

\title{Mel-RoFormer for Vocal Separation and Vocal Melody Transcription}

\oneauthor
 {Ju-Chiang Wang \hspace{0.5cm} Wei-Tsung Lu \hspace{0.5cm} Jitong Chen}
 {ByteDance, San Jose, CA, USA\\
 {\tt\small ju-chiang.wang@bytedance.com}}

\def\authorname{J.-C. Wang, W.-T. Lu, and J. Chen}

\begin{document}

\maketitle
\begin{abstract}
Developing a versatile deep neural network to model music audio is crucial in MIR. This task is challenging due to the intricate spectral variations inherent in music signals, which convey melody, harmonics, and timbres of diverse instruments. In this paper, we introduce Mel-RoFormer, a spectrogram-based model featuring two key designs: a novel Mel-band Projection module at the front-end to enhance the model's capability to capture informative features across multiple frequency bands, and interleaved RoPE Transformers to explicitly model the frequency and time dimensions as two separate sequences.
We apply Mel-RoFormer to tackle two essential MIR tasks: vocal separation and vocal melody transcription, aimed at isolating singing voices from audio mixtures and transcribing their lead melodies, respectively. Despite their shared focus on singing signals, these tasks possess distinct optimization objectives. Instead of training a unified model, we adopt a two-step approach. Initially, we train a vocal separation model, which subsequently serves as a foundation model for fine-tuning for vocal melody transcription.
Through extensive experiments conducted on benchmark datasets, we showcase that our models achieve state-of-the-art performance in both vocal separation and melody transcription tasks, underscoring the efficacy and versatility of Mel-RoFormer in modeling complex music audio signals.
\end{abstract}
\section{Introduction}\label{sec:introduction}

Modeling musical audio signals with deep neural networks (DNNs) for MIR tasks has emerged as a vibrant and promising area of research \cite{dieleman2014end, bittner2022lightweight, choi2016automatic, won2020datadriven, lu2021spectnt}. Most of such DNN models are built upon the spectrogram, a fundamental frequency-time representation of audio signals. Traditional approaches typically treat the spectrogram as a sequence of spectra over time, with the frequency axis representing the feature dimension. However, recent advancements have encompassed explicit modeling of the frequency dimension as a sequence in their architecture designs \cite{zadeh2019wildmix, lu2021spectnt, luo2023music, lu2023multitrack, toyama2023automatic}, recognizing its rich semantic information in music audio signals, including melody, harmonics, and instrument timbres. These architectures have showcased state-of-the-art performance in various MIR tasks such as vocal melody extraction \cite{lu2021spectnt}, section segmentation \cite{wang2022catch}, instrument transcription \cite{lu2023multitrack}, and music source separation \cite{lu2023music}, leveraging the model's ability to discern spectral patterns effectively.

The Transformer architecture \cite{vaswani2017attention} has demonstrated remarkable efficacy not only in Natural Language Processing but also in various MIR tasks, where it excels at modeling sequences to predict high-level musical semantics such as tags, beats, chords, sections, and notes \cite{lu2021spectnt, won2021transformer, lu2023multitrack, toyama2023automatic, wang2022catch, hung2022modeling, yizhi2023mert}. However, its potential to accurately predict low-level audio signals remained uncertain. Lu et al. proposed a novel architecture, called BS-RoFormer \cite{lu2023music}, to tackle the task of music source separation (MSS), which aims to separate audio recordings into musically distinct sources such as vocals, bass, and drums \cite{rafii2018overview, mitsufuji2022music}. Inspired by the Band-Split RNN (BSRNN) model \cite{luo2023music}, BS-RoFormer adopts the interleaved sequence modeling, treating time and frequency dimensions as two separate sequences. Notably, it replaces Recurrent Neural Networks (RNNs) with Transformer encoders, demonstrating exceptional performance. This was evident in its first-place ranking and substantial margin of performance gain over the runner-up in the Music Separation track of the Sound Demixing Challenge 2023 (SDX'23) \cite{fabbro2023sound}.

Another key attribute contributing to the success of BS-RoFormer is the band-split module at the front-end. Traditional Transformer-based models typically rely on a Convolutional Neural Network (CNN) front-end to extract features from the spectrogram for the succeeding Transformer blocks (e.g., \cite{gulati2020conformer, lu2021spectnt, won2021transformer}). However, CNNs are not inherently designed to model two spectral events that are far apart in frequency, which could limit the model to characterize detailed spectral patterns. In contrast, the band-split module divides the frequency dimension into a number of subbands and employs multi-layer perceptrons (MLPs) to directly project the raw subband spectrograms into a sequence of band-wise features for the succeeding Transformer to model it along the frequency axis. From another perspective, the band-split mechanism can be seen as a set of learnable band-pass filters, underscoring the importance of designing an effective band-division scheme.


In this paper, we introduce \emph{Mel-RoFormer}, which is a successor of BS-RoFormer with an enhanced band-division scheme that leverages the Mel-scale \cite{stevens1937scale} to improve the model's generalization ability. The Mel-scale is engineered to mimic the non-linear perception of sound by the human ear, exhibiting higher discrimination at lower frequencies and reduced discrimination at higher frequencies. It has a long history as a reference for designing acoustic features \cite{rabiner2007introduction} such as MFCC and mel-spectrogram in audio signal processing.
We develop the Mel-band mapping based on the Mel-scale, resulting in a band-division that generates overlapping subbands in frequency. In contrast, the band-split scheme in BS-RoFormer is defined empirically and produces non-overlapping subbands.

Mel-RoFormer is applied to address two fundamental MIR tasks: vocal separation and vocal melody transcription, which involve isolating singing voices from audio mixtures and transcribing their lead melodies, respectively. Mel-RoFormer demonstrates superior performance compared to BS-RoFormer and other MSS models in experiments. For vocal melody transcription, we propose a two-step approach instead of training a unified model. We first pretrain a vocal separation model and then fine-tune it for vocal melody transcription. The resulting model achieves state-of-the-art performance across all metrics and exhibits strong robustness in detecting note offsets, which is considered to be the most challenging aspect of the task \cite{molina2014evaluation}. 

Readers can refer to the open-sourced implementation\footnote{\url{https://github.com/lucidrains/BS-RoFormer}} and configurations\footnote{\url{https://github.com/ZFTurbo/Music-Source-Separation-Training}} for Mel-RoFormer and its variants.





\section{Related Work}
\label{sec:related}

To address the capabilities of DNN models that pay attention to modeling the frequency dimension for MIR tasks, SpecTNT is one of early successful attempts. Central to SpecTNT's architecture is the TNT block, where two Transformer encoders are strategically arranged to model along both frequency and time axes. A novel concept introduced by SpecTNT is the Frequency Class Token (FCT), which serves to bridge of the two Transformers, enabling the interchangeability of embeddings across both axes within the TNT block. However, SpecTNT relies on CNNs at the front-end, and the FCT is obtained through aggregating features from the frequency sequence, potentially leading to information loss. Nonetheless, SpecTNT has showcased remarkable performance across various MIR tasks such as beat tracking \cite{hung2022modeling}, chord recognition \cite{lu2021spectnt}, structure segmentation \cite{wang2022catch}, and vocal melody estimation \cite{lu2021spectnt}. Its successor, Perceiver TF \cite{lu2023multitrack}, is designed to enhance efficiency while demonstrating outstanding performance in multitrack instrument/vocal transcription tasks.

Moving on to MSS, a key MIR task that has significantly benefited from DNNs, approaches typically span frequency-domain and time-domain methodologies. The benchmark MUSDB18 dataset \cite{rafii2017musdb18} offers 4-stem sources including vocals, bass, drums, and others, adhering to the definition established by the 2015 Signal Separation Evaluation Campaign (SiSEC) \cite{liutkus20172016}. Frequency-domain approaches rely on spectrogram-based representations as input, leveraging models such as fully connected neural networks \cite{grais2014deep}, CNNs \cite{chandna2017monoaural, kong2021decoupling, jansson2017singing}, and RNNs \cite{uhlich2017improving} to achieve separation. Conversely, time-domain approaches such as Wave-U-Net \cite{stoller2018wave}, ConvTasNet \cite{luo2019conv}, and Demucs \cite{defossez2019music} construct their DNNs directly on waveform inputs. Recently, Hybrid Transformer Demucs (HTDemucs) \cite{rouard2023hybrid} has proposed a novel approach, utilizing a cross-domain Transformer to amalgamate both frequency- and time-domain models, showcasing promising potential in this field. However, none of the mentioned approaches employ Transformers to model the inter-context of frequency and time as two separate sequences.

The output of vocal melody transcription is a sequence of non-overlapping notes, each comprising onset and offset times along with a pitch key, assuming the melody is monophonic. Due to limited training data availability, only a few studies have focused on transcribing note-level outputs from polyphonic music audio, underscoring the significance of pre-training \cite{donahue2022melody} or semi-supervised \cite{kum2022pseudo} techniques. Recently, Wang et al. released a human-annotated dataset comprising 500 Chinese songs \cite{wang2021preparation}, along with a baseline CNN-based model. 
Donahue et al. \cite{donahue2022melody} propose leveraging pre-trained representations from Jukebox \cite{dhariwal2020jukebox} to enhance melody transcription, primarily focusing on lead instruments such as synthesizers, guitars, piano, and vocals. They curate a dataset of 50 hours of melody transcriptions sourced from crowdsourced annotations. However, clarity regarding the quality and identification of vocal melody annotations within songs remains lacking.
In \cite{kum2022pseudo}, a teacher-student training scheme is introduced to leverage pseudo labels derived from fundamental frequency (F0) estimations of vocals.
On the other hand, \cite{hsu2021vocano} presents a system that necessitates a vocal separation as a front-end. In our approach, we employ vocal separation as a pre-trained model for fine-tuning a specialized model, which can be more efficient and task-optimal. 


\begin{figure*}[t]
  \centering
  \centerline{\includegraphics[width=\textwidth]{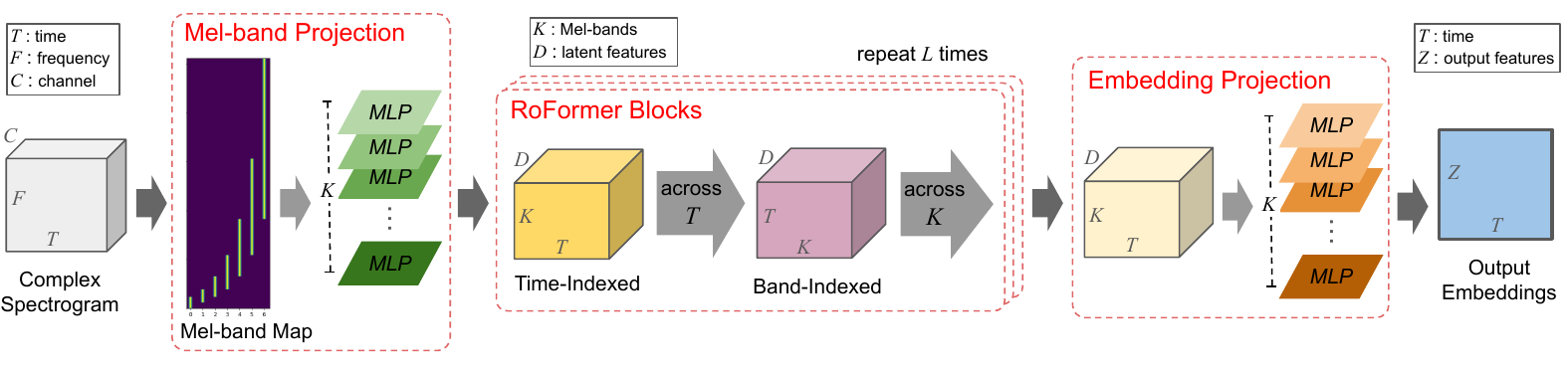}}
  \caption{The diagram of Mel-RoFormer, which is consist of three major modules: Mel-band Projection, RoFormer Blocks, and Embedding Projection. The input is a Complex Spectrogram, and the output is an Embedding tensor, which can be rearranged into the desired shape.}
  \label{fig:melrof_diagram}
\end{figure*}

\section{Model}

Figure \ref{fig:melrof_diagram} illustrates the diagram of Mel-RoFormer, consisting of three major modules: Mel-band Projection, RoFormer blocks, and Embedding Projection. Subsequent subsections will delve into the Mel-band Projection and Embedding Projection modules, while readers can find further details on the RoFormer blocks in \cite{lu2023music}, where they are referred to as ``RoPE Transformer blocks.''

Mel-RoFormer takes the input of a complex spectrogram $X$ with dimensions $(C \times F \times T)$, where $C$, $F$, and $T$ denote the number of channels, frequency bins, and time steps, respectively. This frequency-time representation $X$ is typically obtained via a short-time Fourier transform (STFT), encompassing both real and imaginary parts. In stereo mode, $C$ is defined as $2 \times 2 = 4$, reflecting the presence of two channels for real and imaginary spectrograms.
The output of Mel-RoFormer is denoted by $Y$, with dimensions $(Z \times T)$, where $Z$ and $T$ represent the number of output features and time steps, respectively. One can treat $Y$ as the feature matrix over time. Depending on the downstream tasks, appropriate values for $Z$ can be set, and $Y$ can be rearranged accordingly, as detailed in Section \ref{sec:downstream}.

\subsection{Mel-band Projection Module}
\label{sec:mel-band}

The Mel-band Projection module comprises a frequency-to-Mel-band mapping and a set of $K$ multi-layer perceptrons (MLPs). Each subband of $X$ is denoted as $X_{k}$ with dimensions $(C \times |\mathcal{F}_{k}| \times T)$, where $\mathcal{F}_{k} \in \{0,1,\dots,F-1\}$ represents the indices of frequency bins for the $k$-th subband, and $F$ is the length of frequency bins.

\begin{figure}[t]
  \centering
 \centerline{\includegraphics[width=\columnwidth]{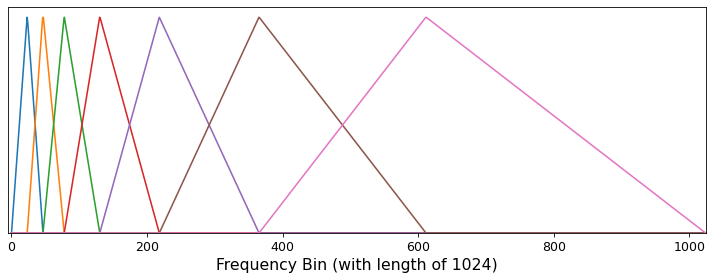}}
  \caption{Illustration of Mel filter-bank with 7 bands. In this example, the length of frequency bins is 1024. Here, the frequency bins from 1 to 46 are encompassed by the 0-th Mel-band (i.e. $\mathcal{F}_0$), those from 24 to 77 are encompassed by the 1-th Mel-band (i.e. $\mathcal{F}_1$), and so forth.}
  \label{fig:mel-7}
\end{figure}

The frequency-to-Mel-band mapping, represented as $\{\mathcal{F}_{k}\}_{k=0}^{K-1}$, stems from the Mel filter-bank, as depicted in Fig. \ref{fig:mel-7}, where triangular-shaped filters are centered at different Mel frequencies on the Mel-scale. The indices of non-zero values of a filter correspond to the frequency bins of the respective Mel-band. Mathematically, the Mel-scale follows a quasi-logarithmic function of acoustic frequency, ensuring that perceptually similar pitch intervals (e.g., octaves) possess equal width across the entire audible range.
The width of a Mel-band (i.e., $|\mathcal{F}_{k}|$) is twice the distance between its center and the center of the preceding Mel-band. Consequently, the latter half of a Mel-band overlaps with its subsequent Mel-band, and so forth, until reaching the final Mel-band. In contrast, due to the band-split design in BS-RoFormer \cite{lu2023music}, the frequency ranges of different subbands do not overlap. The Mel-band Map depicted in Fig. \ref{fig:melrof_diagram} exemplifies the frequency-to-Mel-band mapping, illustrating a binary relationship between 1024 frequency bins and 7 Mel-bands.

The input $X_k$ is rearranged into a shape of $(C|F_{k}| \times T)$ for the MLP layer.  The $k$-th MLP, denoted as $\Lambda_k$, comprises an RMSNorm layer \cite{zhang2019root} followed by a linear layer. The linear layer transforms from $C |F_{k}|$ dimensions to $D$ dimensions, where $D$ is the number of latent features. 
The resulting outputs $\{\Lambda_k(X_k)\}_{k=0}^{K-1}$ are stacked to form a shape of $(D \times K \times T)$, serving as input to the subsequent RoFormer blocks.
The Mel-band Projection module can be conceptualized as a learnable Mel filter-bank, with the MLP layer functioning as the mechanism to learn the filters. This grants the model greater flexibility in determining the optimal shapes for different filters, without being confined to predefined filter designs such as triangle-shaped ones (see Fig. \ref{fig:mel-7}).


\subsection{RoFormer Blocks}
\label{sec:roformer-blocks}

The RoForMer blocks consist of a stack of $L$ interleaved RoPE Transformer encoders \cite{su2021roformer}. The interleaved sequence modeling processes the data across time ($T$) and subband ($K$) dimensions alternately. In BS-RoFormer \cite{lu2023music}, the authors observed that Rotary Position Encoding (RoPE) played a crucial role in enhancing Transformer performance compared to using traditional absolute position encoding. It is suggested that RoPE aids in preserving the positional information within the sequence, making it invariant to repetitive processes of rearrangement.

Specifically, the data is first rearranged into a time-indexed shape of $(DK \times T)$, allowing modeling across time. Subsequently, it is rearranged into a band-indexed shape of $(DT \times K)$, facilitating modeling across subbands. The former step treats the data as a time sequence, while the latter treats it as a subband sequence. By repeating this process, information from different time steps and subbands becomes interchangeable, thereby enhancing the model's ability to generalize.

\subsection{Embedding Projection Module}
\label{sec:embed-mlp}

The Embedding Projection module plays a crucial role in generating suitable embeddings necessary for various downstream tasks. It has been observed that utilizing MLPs can lead to more effective mask estimation compared to using plain linear layers in source separation tasks \cite{li2022use}. Our preliminary investigation also suggests that removing this module can result in unstable training. 

This module comprises $K$ individual MLPs, denoted as $\Phi_k$, each containing an RMSNorm layer, a linear layer followed by a Tanh activation, and another linear layer followed by a gated linear unit (GLU) layer \cite{dauphin2017language}. The first linear layer transforms from $D$ to $4D$ dimensions, while the subsequent linear layer with GLU transforms from $4D$ to a desired length $Z_k$. Here, $Z_k$ specifies the number of output features for $\Phi_k$ based on its specific purpose, as elaborated in Section \ref{sec:downstream}. All the MLP outputs are concatenated along the feature dimension, resulting in a final output shape of $(Z \times T)$, where $Z=\sum_k Z_k$.

\section{Downstream Tasks}
\label{sec:downstream}

This section details the application of Mel-RoFormer tailored for vocal separation and vocal melody transcription.

\subsection{Vocal Separation}
\label{sec:vocal-separation}

For vocal separation, the Embedding Projection estimates the mask for the complex spectrogram. Each MLP $\Phi_k$ is designed to align its output shape with that of the corresponding input Mel-band complex spectrogram, with $Z_k$ set as $C|\mathcal{F}_k|$. The output $Y$ is rearranged into 
\begin{equation} 
\label{eq:estimate_mask}
 \left[ \hat{\Phi}_0, \hat{\Phi}_1, \dots, \hat{\Phi}_{(K-1)}  \right],
\end{equation}
with dimensions $(C \times \hat{Z} \times T)$, where $\hat{\Phi}_k$ with a shape of $(|\mathcal{F}_k| \times T)$ is the output corresponding to the $k$-th Mel-band, and $\hat{Z} = \sum_k |\mathcal{F}_k|$.
Because adjacent Mel-bands overlap, the estimated mask values of the overlapping frequency bins are averaged:
\begin{equation} 
\label{eq:avg_mask}
\hat{M}[c, f, t] = \frac{1}{S_f} \sum_{k} \hat{\Phi}_k[c, f, t] , 
\end{equation}
where $\hat{M}$ represents the estimated mask, $c$, $f$, and $t$ are the indices of channel, frequency bin, and time step, respectively, and $S_f$ is the count of the overlapping frequency bins. 
The estimated mask $\hat{M}$ has the same shape $(C \times F \times T)$ as that of the input $X$, encompassing both the real and imaginary parts of the complex spectrogram.

We utilize complex Ideal Ratio Masks (cIRMs) \cite{wang2018supervised} as our optimization goal for the vocal separation model. The estimated mask $\hat{M}$ derived from Embedding Projection can serve as the cIRMs. The separated complex spectrogram $\hat{Y}$ is obtained by element-wise multiplication of the cIRM with the input complex spectrogram: $ \hat{Y} = \hat{M} \odot X $. Subsequently, an inverse STFT (iSTFT) is applied to $ \hat{Y} $ to reconstruct the separated signal $\hat{y}$ in the time-domain. 

Let $\psi$ denote the target time-domain signal, and $\Psi^{(w, r)}$ denote the corresponding complex Spectrogram using a window size $w$ and time-resolution $r$ for STFT. We employ the mean absolute error (MAE) loss to train the cIRMs $\hat{M}$. Specifically, the objective loss encompasses both the time-domain MAE and the multi-resolution complex spectrogram MAE \cite{guso2022loss}:
\begin{equation}
\label{eq:loss}
\mathcal{L} = || \psi - \hat{y} || + \sum_{w \in W, r \in R} || \Psi^{(w, r)} - \hat{Y}^{(w, r)}||,
\end{equation}
where the configurations for multi-resolution STFTs cover 5 window sizes with $W$ = $\{4096, 2048, 1024, 512, 256\}$, and 2 resolutions with $R$ = $\{100, 300\}$ frames per second.

\subsection{Vocal Melody Transcription}
\label{sec:melody-trans}

Instead of starting training from scratch, we utilize a pre-trained vocal separation Mel-RoFormer and fine-tune it for vocal melody transcription. Given that the Embedding Projection module in the pre-trained vocal separation model functions as a mask estimator, it might inherently possess biases towards signal-level semantics. Therefore, we opt to replace the pre-trained Embedding Projection with a newly initialized one, with a modification specifically on the output dimension. To ensure equal contribution from each Mel-band to the feature dimension, we set a uniform value of $64$ for all $Z_k$'s. The resulting embeddings take the shape of $(64K \times T)$, with a $64K$-dimensional feature vector for each time step.

We adopt the "onsets and frames" approach \cite{hawthorne2017onsets}, employing two frame-wise predictors: an onset predictor and a frame predictor, both of which receive embeddings from Mel-RoFormer. The onset predictor identifies the onset event of a pitched note, while the frame predictor determines the continuation of a pitched note. This design facilitates a post-processing method where the initiation of a new note is determined only if the onset predictor indicates the start of a pitch, and simultaneously, the frame predictor confirms the presence of an onset for that pitch within the succeeding frames. The onset and frame predictors operate at a time-resolution of 50 frames per second. In cases where the Mel-RoFormer embeddings do not match this time-resolution, we employ 1-D adaptive average pooling over the time dimension.

We employ an MLP layer for the onset predictor, consisting of a linear layer followed by a Rectified Linear Unit (ReLU), dropout with a rate of $0.5$, and another linear layer. The linear layer has $512$ hidden channels, and the output dimension is set to 60, representing 60 supported pitches. For the frame predictor, a single linear layer is utilized, outputting 61 pitch classes, with one indicating non-pitch. Binary cross-entropy losses of the two predictors are summed as the final loss to train the entire model. The thresholds for the onset and frame predictors are set at $0.45$ and $0.25$, respectively.

\section{Experiment}
\label{sec:exp}

Our experiments cover vocal separation and vocal melody transcription. In the vocal separation evaluation, we train and test two types of models: the first on 44.1kHz stereo audio recordings, and the second on 24kHz mono audio recordings. Next, we utilize the model trained on 24kHz mono audio recordings as the pre-trained model for fine-tuning in the vocal melody transcription evaluation.

\subsection{Datasets}

\begin{table}[t]
\centering
\resizebox{\columnwidth}{!}{%
\begin{tabular}{lccl}
 \toprule
 Dataset  & Task & Songs & Split \\
 \midrule
MUSDB18HQ~\cite{rafii2017musdb18} & sepa & 150 & train 100, test 50 \\
MoisesDB~\cite{pereira2023moisesdb} & sepa & 240 & train 200, val 40 \\
MIR-ST500~\cite{wang2021preparation} & trans & 500 & train 330, val 37, test 98 \\
POP909~\cite{wang2020pop909} & trans & 909 & train 750, val 50, test 109 \\
In-House & sepa & 1533 & train 1433, val 100 \\
 \bottomrule
 \end{tabular}}
\caption{Summary of the datasets used in this study. Abbreviations: `sepa': vocal separation, `trans': vocal melody transcription, 'val': validation. }
\label{table:dataset}
\end{table}

Table \ref{table:dataset} overviews the datasets used in this study. We use four public datasets for evaluation. The data splitting adheres to the official guidelines of each dataset, except for POP909, where songs with IDs ranging from 801 to 909 are reserved for testing. The `Split' column of Table \ref{table:dataset} indicates the numbers of songs allocated for training, validation, and testing. All data for separation tasks are stereo recordings with a sampling rate of 44.1kHz, and stem-level recordings are pre-mixed into four stems: vocals, bass, drums, and other. The audio of transcription data was resampled to mono with a 24kHz sampling rate to follow the conventional setting \cite{hsu2021vocano}. Although access to some songs in MIR-ST500 was restricted, our test set, comprising 98 songs, closely resembles the original setting of 100 songs.

\subsection{Configuration for Vocal Separation}

To obtain the frequency-to-Mel-band mapping, we employ the Mel filter-bank implementation in librosa \cite{mcfee2015librosa}, which emulates the behavior of the function in MATLAB Auditory Toolbox \cite{slaney1998auditory}. By using \texttt{librosa.filters.mel}, we acquire the mapping matrix comprising a triangle filter for each Mel-band. Subsequently, we binarize this matrix by setting all non-zero values to 1, thereby discarding the triangle filters. This process yields the Mel-band Map depicted in Figure \ref{fig:melrof_diagram}.

We follow the method outlined in \cite{lu2023music} for performing random remixing data augmentation. This strategy involves cross-dataset stem-level combination, resulting in a significantly larger number of examples than the original size of the datasets combined.

Tree evaluation scenarios are considered: \textcircled{a} \emph{musdb18-only}: train a 44.1kHz stereo model only on the MUSDB18-HQ training set; \textcircled{{b}} \emph{all-data}: all additional data, including MUSDB18HQ, MoisesDB, and In-House, are used to train a 44.1kHz stereo model; \textcircled{{c}} \emph{musdb18+moisesdb}: MUSDB18HQ and MoisesDB are resampled to train a 24kHz mono model, serving as the pre-trained model for fine-tuning for the melody transcription task.

Our main baseline is BS-RoFormer \cite{lu2023music}. For scenarios \textcircled{{a}} and \textcircled{{b}}, we set the parameters as follows: $T$=800 (8-second chunk), $K$=60, $D$=384, $L$=12, and a window size of 2048 and a hop size of 441 for STFT. In scenario \textcircled{{c}}, two models are trained: \emph{24k-small} and \emph{24k-large}. The small model uses $T$=300 (6-second chunk), $K$=32, $D$=128, $L$=12; while the large model uses $T$=300 (6-second chunk), $K$=32, $D$=256, $L$=24. Both models adopt a window size of 1024 and a hop size of 480 for STFT. All the separation models use the ``overlap \& average'' deframing method \cite{lu2023music} with a hop of half a chunk. These above mentioned settings remain consistent between Mel-RoFormer and BS-RoFormer. 

For training, we utilize the AdamW optimizer \cite{loshchilov2017decoupled} with a learning rate (LR) of 0.0005. The LR is reduced by 10\% every 40k steps. To optimize GPU memory usage, we employ flash-attention \cite{dao2023flashattention} and mixed precision. Specifically, the STFT and iSTFT modules use FP32, while all other components use FP16.
Regarding hardware configurations, we employ different setups for each scenario. In scenario \textcircled{a}, 8 Nvidia A-100-80GB GPUs with batch\_size=64 are used, and the training stopped at 400K steps (\textasciitilde 40 days). For scenario \textcircled{b}, 16 Nvidia A100-80GB GPUs with batch\_size=128 are utilized, and the training halted at 1M steps (\textasciitilde 93 days). In scenario \textcircled{c}, 16 Nvidia V100-32GB GPUs with batch\_size=96 are used, and the training stopped at 500K steps (\textasciitilde 31 days).

The reason to use a large number of training steps is driven by the continuous improvement observed in the model's performance, coupled with the absence of overfitting. This can be attributed to two main factors: the effect of the random remixing augmentation and the inherent capability of the model itself. These factors contribute to the model's ability to continuously learn and adapt to the training data, resulting in sustained performance improvements without encountering overfitting issues.

\begin{table}[t]
\begin{tabular}{lcc}
 \toprule
 Model & Vocals  & \# Param \\
 \midrule
 HDemucs \cite{defossez2021hybrid}$^\dag$ & 8.04 & - \\
 Sparse HT Demucs \cite{rouard2023hybrid}$^\dag$ & 9.37 & - \\
 BSRNN \cite{luo2023music}$^\dag$ & 10.01 & - \\
 TFC-TDF-UNet-V3 \cite{kim2023sound}$^\dag$ & 9.59 & - \\
 BS-RoFormer \textcircled{{a}}  & 11.49 & 93.4M \\
 Mel-RoFormer \textcircled{{a}} & 12.08 & 105M\\
 BS-RoFormer \textcircled{{b}}$^\dag$  & 12.82& 93.4M \\
 Mel-RoFormer \textcircled{{b}}$^\dag$ & \textbf{13.29} & 105M \\ 
 \toprule
 \multicolumn{3}{c}{\emph{Tested with resampled 24kHz mono audio}} \\
 \midrule
 BS-RoFormer (24k-small) \textcircled{{c}}  & 10.56 & 8.0M \\
 Mel-RoFormer (24k-small) \textcircled{{c}} & 11.01 & 9.1M \\ 
 BS-RoFormer (24k-large) \textcircled{{c}}  & 12.19 & 48.4M \\
 Mel-RoFormer (24k-large) \textcircled{{c}} & \textbf{12.69} & 50.7M \\ 
 \bottomrule
 \end{tabular}
\small{Symbol $\dag$ indicates models trained with extra data.}  \\
\small{Symbols \textcircled{a}, \textcircled{b}, and \textcircled{c} indicate the three evaluation scenarios.}  
\caption{Result (in SDR) on MUSDB18HQ test set.}
\label{table:mss_result}
\end{table}

\subsection{Result for Vocal Separation}

Table \ref{table:mss_result} presents the results, with the signal-to-distortion ratio (SDR) values \cite{vincent2006performance} computed by \texttt{museval} \cite{SiSEC18} as the evaluation metric. The median SDR across the median SDRs over all 1-second chunks of each test song is reported, following prior conventions. Several representative existing models are included for comparison.

From the result, we see that Mel-RoFormer achieve state-of-the-art performance. It is evident that Mel-band Projection significantly enhances vocal separation performance, leading to a consistent improvement over BS-RoFormer, with an average gain of 0.5 dB across all scenarios. This showcases the effectiveness of the Mel-band mapping scheme in capturing human voices. Qualitative analysis indicates that Mel-RoFormer produces smoother vocal sounds with more consistent loudness. On the other hand, the 24kHz mono model also performs admirably, which bodes well for downstream tasks like vocal melody transcription, as they do not necessitate high-quality audio with high sampling rates. Furthermore, the smaller model with 9.1M parameters achieves over 11 dB, demonstrating its potential for resource-constrained environments.

\begin{table}[t]
\begin{tabular}{lcccc}
 \toprule
 \small{Model} & \small{\#Param} & \small{COn} & \small{COnP} & \small{COnPOff} \\
 \midrule
 Efficient-b1 \cite{wang2021preparation} & - & .754 & .666 & .458 \\
 JDC$_{\text{note}}$ \cite{kum2022pseudo} & - & .762 & .697 & .422 \\
 A-VST \cite{gu2023deep} & - & .783 & .707 & .538 \\
 Perceiver TF \cite{lu2023multitrack} & - & - & .777 & - \\
 MERT \cite{yizhi2023mert} \textcircled{d} & 324M & .775 & .751 & .530 \\
 SpecTNT \cite{lu2021spectnt} \textcircled{d}  & 8.4M & .801 & .778 & .550\\
 Mel-RoF-small \textcircled{d} & 14.5M & .807 & .786 & .609 \\
 Mel-RoF-large \textcircled{d} & 64.6M & \textbf{.819} & \textbf{.798} & \textbf{.625} \\ 
 Mel-RoF-small \textcircled{f} & 14.5M & .780 & .765 & .574 \\
 Mel-RoF-large \textcircled{f} & 64.6M & .790 & .776 & .594 \\ 
 \bottomrule
 \end{tabular}
\centering
\small{Symbols \textcircled{d}, \textcircled{e}, and \textcircled{f} indicate three evaluation scenarios.}~~~~  \\
\caption{Model comparison on MIR-ST500 test set.}
\label{table:mirst500}
\end{table}

\begin{table}[t]
\begin{tabular}{lcccc}
 \toprule
 Model & COn & COnP & COnPOff \\
 \midrule
 MERT \cite{yizhi2023mert} \textcircled{e} & .745 & .697 & .315 \\
 SpecTNT \cite{lu2021spectnt} \textcircled{e} & .797 & .775 & .371\\
 Mel-RoF-small \textcircled{e} & .831 & .805 & .398 \\
 Mel-RoF-large \textcircled{e} & \textbf{.869} & \textbf{.842} & .486 \\ 
 Mel-RoF-small \textcircled{f} & .833 & .808 & .405 \\
 Mel-RoF-large \textcircled{f} & .864 & .839 & \textbf{.494} \\ 
 \bottomrule
 \end{tabular}
\centering
\caption{Model comparison on POP909 test set. Evaluated with a time tolerance of 80 ms.}
\label{table:pop909}
\end{table}

\subsection{Configuration for Vocal Melody Transcription}

Three evaluation scenarios are studied: \textcircled{d} trained on MIR-ST500; \textcircled{e} trained on POP909, and \textcircled{f} trained on a combination of MIR-ST500 and POP909. With $T$=300 and $K$=32 for Mel-RoFormer, the resulting embedding matrix for the onset and frame predictors has a shape of $(2048 \times 300)$, representing a 6-second input with a frame rate of 50Hz. We also consider a variant that is trained from scratch without a pre-trained model.

For fine-tuning, we use the AdamW optimizer with a LR of 0.001 for the onset and frame predictors and 0.0001 for the Mel-RoFormer module. The LR is reduced by 10\% with a patience of 15 epochs, where an epoch comprises 100 steps. The best model is selected based on validation performance for testing. 

For baselines, we implement SpecTNT \cite{lu2021spectnt} and MERT \cite{yizhi2023mert}. SpecTNT features 5-layer TNT blocks and is trained with the random mixing augmentation method outlined in \cite{lu2023multitrack}. For MERT, we use the pre-trained weights of ``MERT-v1-330M,'' where the model accepts 5-second input with a 24kHz audio sampling rate and produces embeddings with a shape of $(1024 \times 375)$. All mentioned models are trained or fine-tuned using 8 Nvidia V100-32GB GPUs.

Evaluation metrics include the F-measures of Correct Onset (COn), Correct Onset and Pitch (COnP), and Correct Onset, Pitch, and Offset (COnPOff). These metrics are computed using \texttt{mir\_eval} \cite{raffel2014mir_eval}, with a pitch tolerance of 50 cents and a time tolerance of 50 ms. For POP909, we adjust the time tolerance to 80ms due to the less precise nature of note onsets and offsets in its labeling method \cite{wang2020pop909}.

\subsection{Result for Vocal Melody Transcription}


Tables \ref{table:mirst500} and \ref{table:pop909} display the results on MIR-ST500 and POP909, respectively, featuring a comparison with various existing models. 
It is worth noting that fine-tuning a pre-trained model typically converges in fewer than 15K steps, while training from scratch requires significantly more steps (e.g., 50k steps) and yields significantly inferior performance compared to fine-tuned models.

Several key observations emerge from the results. Firstly, our proposed model (Mel-RoF-large and Mel-RoF-small) achieves state-of-the-art performance across all metrics on both MIR-ST500 and POP909, showcasing its effectiveness. Particularly noteworthy is the substantial performance improvement in COnPOff over the baselines (e.g., a 7.5 percentage point increase in Mel-RoF-large compared to SpecTNT), highlighting the robustness of Mel-RoFormer in accurately determining full notes, including onset, pitch, and offset. This can be attributed to its capability to extract clean singing voices, thereby minimizing the influence of irrelevant instruments. Comparing our models to MERT underscores the superiority of Mel-RoFormer, owing to its architectural design and pre-training with a relevant task. This emphasizes the importance of explicitly modeling frequency with Transformers and suggests that the separation task can be a valuable objective when training a foundation model. 

Cross-comparing MIR-ST500 and POP909, we note that annotations are relatively more consistent in MIR-ST500. In contrast, POP909 exhibits errors primarily at note offsets, along with global time shifts in several songs. Consequently, we accept a larger time tolerance of 80 ms in the evaluation scenario. Particularly, in scenario \textcircled{f}, training with combined datasets improves test performance on POP909 but degrades that of MIR-ST500, consistent with our observations about labeling quality.

\section{Conclusion}

We have presented the Mel-RoFormer model, which integrates the Mel-band Projection scheme to enhance its ability to model musical signals effectively. Our experiments have shown highly promising results in vocal separation and vocal melody transcription. These findings suggest the potential of Mel-RoFormer as a foundation model for various other MIR tasks, including chord recognition and multi-instrument transcription \cite{cheuk2023jointist}. 

\bibliography{mss_transcription}
\end{document}